\documentclass[aps,superscriptaddress,unsortedaddress,showpacs]{revtex4}
\usepackage{amsmath}
\usepackage{amssymb}
\usepackage{graphics}
\usepackage{epsfig}
\usepackage{color}

\begin{document}

\title{Quasi-energies, parametric resonances, and stability limits in
ac-driven $\mathcal{PT}$-symmetric systems }
\author{Jennie D'Ambroise}
\affiliation{Department of Mathematics, Amherst College, Amherst, MA 01002-5000, USA}
\author{Boris A. Malomed}
\affiliation{Department of Physical Electronics, School of Electrical Engineering,
Faculty of Engineering, Tel Aviv University, Tel Aviv 69978, Israel}
\author{P.G. Kevrekidis}
\affiliation{Department of Mathematics and Statistics, University of Massachusetts,
Amherst, MA 01003-9305, USA}

\begin{abstract}
We introduce a simple model for implementing the concepts of quasi-energy
and parametric resonances (PRs) in systems with the $\mathcal{PT}$ symmetry,
i.e., a pair of coupled and mutually balanced gain and loss elements. The
parametric (ac) forcing is applied through periodic modulation of the
coefficient accounting for the coupling of the two degrees of freedom. The
system may be realized in optics as a dual-core waveguide with the gain and
loss applied to different cores, and the thickness of the gap between them
subject to a periodic modulation. The onset and development of the
parametric instability for a small forcing amplitude ($V_{1}$) is studied in
an analytical form. The full dynamical chart of the system is generated by
systematic simulations. At sufficiently large values of the forcing
frequency, $\omega $, tongues of the parametric instability originate, with
the increase of $V_{1}$, as predicted by the analysis. However, the tongues
following further increase of $V_{1}$ feature a pattern drastically
different from that in usual (non-$\mathcal{PT}$)\ parametrically driven
systems: instead of bending down to larger values of the dc coupling
constant, $V_{0}$, they maintain a direction parallel to the $V_{1}$ axis.
The system of the parallel tongues gets dense with the decrease of $\omega $%
, merging into a complex small-scale structure of alternating regions of
stability and instability. The cases of $\omega \rightarrow 0$ and $\omega
\rightarrow \infty $ are studied analytically by means of the adiabatic and
averaging approximation, respectively. The cubic nonlinearity, if added to
the system, alters the picture, destabilizing many originally robust
dynamical regimes, and stabilizing some which were unstable.
\end{abstract}

\maketitle


\textbf{Recently, a great deal of interest has been drawn to systems
featuring the }$\mathcal{PT}$\textbf{\ (parity-time) symmetry. Originally,
they were introduced as quantum systems with spatially separated and
symmetrically placed linear gain and loss. A fundamental property of
non-Hermitian }$\mathcal{PT}$\textbf{-symmetric Hamiltonians is the fact
that their spectra remain purely real, like in the case of the usual
Hermitian Hamiltonians, provided that the strength of the non-Hermitian part
of the Hamiltonian, }$\gamma $, \textbf{does not exceed a certain critical
value, }$\gamma _{\mathrm{cr}}$\textbf{, at which the }$\mathcal{PT}$\textbf{\
symmetry breaks down. Such linear systems were recently implemented
experimentally in optics, due to the fact that the paraxial propagation equation
for electromagnetic waves can emulate the quantum-mechanical Schr\"{o}dinger
equation. In particular, the }$\mathcal{PT}$\textbf{-symmetric quantum
system may be emulated by waveguides with spatially separated mutually
balanced gain and loss elements. Additional interest to the optical
realizations of the }$\mathcal{PT}$\textbf{\ symmetry has been drawn by the
possibility to implement this setting in nonlinear waveguides. Unlike the
ordinary models of nonlinear dissipative systems, where stable modes exist
as isolated attractors, in }$\mathcal{PT}$\textbf{-symmetric systems they
emerge in continuous families, similar to what is commonly known about
conservative nonlinear systems. However, the increase of the gain-loss
coefficient, }$\gamma $\textbf{, leads to the shrinkage of existence and
stability regions for the }$\mathcal{PT}$\textbf{-symmetric modes, which
completely vanish at some }$\gamma =\gamma _{\mathrm{cr}}$\textbf{. Especially
convenient for the study of such solutions are models of \textit{dual-core
couplers}, with the balanced gain and loss applied to different cores, and
the cubic nonlinearity acting in both. The objective of the present work is
to extend the well-known concepts of quasi-energy and parametric resonances
(PRs) to }$\mathcal{PT}$\textbf{-symmetric systems. In terms of the coupler model, the
parametric drive (with ac amplitude }$V_{1}$ \textbf{and dc amplitude} $%
V_{0} $\textbf{) can be implemented by making the thickness of the gap
between the two cores in the coupler a periodic function of the evolution
variable (propagation distance, in terms of optics), with frequency }$\omega
$\textbf{. Using a combination of analytical and numerical methods, we
produce the full dynamical chart for both linear and nonlinear versions of
the system in the plane of }$\left( V_{1},V_{0}\right) $\textbf{\ for
different values of } $\omega $\textbf{. The charts feature tongues of
parametric instability, whose shape has some significant differences from
that predicted by the classical theory of the PR in conservative systems. In
addition, the cases of small and large }$\omega $\textbf{\ are studied by
means of the adiabatic and averaging approximations, respectively.}

\section{Introduction and the model}

A commonly known principle of quantum mechanics is that operators
representing physical variables, including the Hamiltonian, must be
Hermitian, to guarantee the reality of the respective spectra. Recently, it
was found that the class of physically acceptable Hamiltonians may be
extended by admitting operators with anti-Hermitian [dissipative and
antidissipative (gain)] parts, if they commute with the $\mathcal{PT}$
(parity-time) transformation \cite{Bender_review}. The spectrum of such a $%
\mathcal{PT}$-invariant Hamiltonian may be purely real, provided that the
strength of the anti-Hermitian part, $\gamma $, does not exceeds a critical
value, $\gamma _{\mathrm{cr}}$, at which complex eigenvalues emerge. In the
simplest form, a $\mathcal{PT}$-invariant Hamiltonian contains a complex
potential in which the real part is an even function of coordinates, while
the imaginary part is odd.

While in the quantum theory this concept remains a rather abstract one, the
similarity of the quantum-mechanical Schr\"{o}dinger equation to the
paraxial propagation equation in optics has suggested a possibility to
emulate the $\mathcal{PT}$-symmetry in optical waveguides with symmetrically
placed mutually balanced gain and loss elements. This implementation was
proposed theoretically~\cite{Muga} and then demonstrated in experiments \cite%
{experiment}. Recently, another experimental realization of the $\mathcal{PT}
$\ symmetry was reported in a system of coupled electronic oscillators \cite%
{Kottos}. There is a possibility to realize such a setting in Bose-Einstein
condensates \cite{BEC} too.

These possibilities have drawn a great deal of interest to the analysis of
various $\mathcal{PT}$-symmetric dynamical models \cite%
{special-issues,review}. In particular, the natural occurrence of the
self-focusing in optical media stimulated the study of nonlinear effects in
such systems. The interplay of the paraxial diffraction, Kerr nonlinearity,
and a spatially periodic complex potential, which represents the $\mathcal{PT%
}$-symmetric part of the system, gives rise to bright solitons \cite%
{Musslimani2008}, whose stability was rigorously analyzed in Ref.~\cite{Yang}%
. Dark solitons have been predicted too in this context, assuming the
self-defocusing Kerr nonlinearity \cite{dark}. In addition, bright solitons
have been predicted in $\mathcal{PT}$-symmetric systems with the quadratic
(second-harmonic-generating) nonlinearity \cite{chi2}. Systems of another
type, which readily maintain $\mathcal{PT}$-symmetric solitons, and actually
make it possible to obtain such solutions and their stability conditions in
an exact analytical form, are provided by models of linearly-coupled
dual-core waveguides, with the balanced gain and loss applied to the two
cores, and the intrinsic Kerr nonlinearity present in each one \cite{dual}-%
\cite{dark-coupler}.

Further, nonlinear modes were studied in many discrete $\mathcal{PT}$%
-symmetric systems with the cubic nonlinearity. These include linear \cite%
{discrete} and circular \cite{circular} chains of coupled $\mathcal{PT}$
elements and aggregates of coupled $\mathcal{PT}$-symmetric oligomers
(dimers \cite{Hadi}, quadrimers, etc.)~\cite{KPZ}. General features of the
dynamics in $\mathcal{PT}$-symmetric lattices of single-component nonlinear $%
\mathcal{PT}$-symmetric lattices have been recently considered in~Ref. \cite%
{pelinov}. Related to these models are systems featuring discrete $\mathcal{%
PT}$-balanced elements embedded into continuous nonlinear conservative media
\cite{Wunner,Thaw}, which, in particular, allow one to find exact solutions
for solitons pinned to the $\mathcal{PT}$-\textit{symmetric dipole} \cite%
{Thaw}.

The inclusion of the Kerr nonlinearity into the conservative part of the
system suggests that its gain-and-loss part can be made nonlinear too, by
adopting mutually balanced cubic gain and loss terms ~\cite%
{AKKZ,Canberra,rudy}. Effects of combined linear and nonlinear $\mathcal{PT}$%
~terms on the existence and stability of solitons were studied in \cite%
{combined}.

In the usual nonlinear dissipative systems, stable dynamical regimes exist
as isolated solutions (\textit{attractors}), such as limit cycles or strange
attractors \cite{SA} in discrete settings, or dissipative solitons in
dissipative media \cite{PhysicaD}. On the contrary to that, in $\mathcal{PT}$%
-symmetric systems stable modes form continuous families, similar to the
generic situation in conservative systems. However, the increase of the
gain-loss coefficient ($\gamma $) in the $\mathcal{PT}$-symmetric system
leads to shrinkage of existence and stability domains for the families,
which eventually vanish at $\gamma =\gamma _{\mathrm{cr}}$.

In many cases, quantum systems are driven by time-periodic (ac) external
forces. In that case, the eigenstates are characterized by particular values
of the quasi-energy, defined as per the Floquet theorem, see, e.g., Ref.
\cite{Floquet}. A related concept is the parametric resonance (PR) in linear
and nonlinear systems subject to the action of the parametric drive \cite{LL}%
. In the context of $\mathcal{PT}$-symmetric systems, it is interesting to
analyze the realization of the quasi-energy and PRs in ac-driven systems,
both linear and nonlinear ones, including mutually balanced gain and loss
terms. This analysis is the subject of the present work. We perform it in
the framework of a basic model with two degrees of freedom (a \textit{dimer}%
), represented by complex amplitudes $\psi _{A}$ and $\psi _{B}$, and the
coupling containing an ac-modulated term, which represents the parametric
drive:%
\begin{eqnarray}
i\frac{d}{dt}\psi _{A} &=&i\psi _{A}+\left[ V_{0}+2V_{1}\cos \left( \omega
t\right) \right] \psi _{B}+\chi |\psi _{A}|^{2}\psi _{A},  \notag \\
i\frac{d}{dt}\psi _{B} &=&-i\psi _{B}+\left[ V_{0}+2V_{1}\cos \left( \omega
t\right) \right] \psi _{A}+\chi |\psi _{B}|^{2}\psi _{B}.  \label{v2}
\end{eqnarray}%
Here the equal gain and loss coefficients in the equations for $\psi _{A}$
and $\psi _{B}$ are scaled to be $\gamma \equiv 1$, $V_{0}$ and $V_{1}$ are
strengths of the constant (dc) and ac-modulated couplings, $\omega $ is the
modulation frequency, and $\chi $ is the nonlinearity coefficient, which may
be taken as $\chi =0$ or $\chi \equiv 1$ in the linear and nonlinear
versions of the system. Note that by the complex conjugation and replacing $%
\psi _{A},\psi _{B}$ $\rightarrow $ $-\psi _{A}^{\ast },\psi _{B}^{\ast }$
Eqs. (\ref{v2}) with $\chi =\pm 1$ are transformed into each other,
therefore we do not consider $\chi =-1$. Also, by an obvious transformation $%
 \psi _{A},\psi _{B} \rightarrow  \psi _{A},-\psi
_{B} $, it is possible to reverse the sign of $V_{0}$, and shift $%
t\rightarrow t+\pi /\omega $ can independently reverse the sign of $V_{1}$,
therefore we can fix both $V_{0}$ and $V_{1}$ to be positive.

According to the concept of the quasi-energy, quasi-stationary solutions to
Eqs. (\ref{v2}) for $\omega \neq 0$ are sought for as%
\begin{equation}
\psi _{A}(t)=\exp \left( -iEt\right) \Psi _{A}(t),~\psi _{B}(t)=\exp \left(
-iEt\right) \Psi _{B}(t)  \label{quasi}
\end{equation}%
where $\Psi _{A,B}(t)$ are periodic functions with period $2\pi /\omega $,
and the quasi-energy $E$ belongs to interval $0<E<2\pi /\omega $. A
corollary of Eqs. (\ref{v2}) is the balance equation for the total power in
the presence of the mutually symmetric gain and loss~\cite{pelinov}%
\begin{equation}
\frac{d}{dt}\left( |\psi _{A}|^{2}+|\psi _{B}|^{2}\right) =2\left( |\psi
_{A}|^{2}-|\psi _{B}|^{2}\right) .  \label{D}
\end{equation}

This system can be implemented as a model of a dual-core optical waveguide
operating in the CW (continuous-wave) regime, so that Eqs. (\ref{v2}) do not
include terms accounting for the transverse diffraction or group-velocity
dispersion, in the spatial-domain and temporal-domain realizations,
respectively. Variable $t$ is actually the propagation distance, and the ac
modulation represents periodic variation of the thickness of the gap
separating the cores \cite{Sydney}. As suggested in Refs. \cite{dual}, terms
$\pm i\psi _{A,B}$ in Eqs. (\ref{v2}) represent the gain and loss applied to
the two cores. A similar system, without the ac-modulating terms, but with
the addition of cubic gain and loss terms, was proposed as a model of a $%
\mathcal{PT}$-symmetric dimer in Ref. \cite{Canberra}.

The main issue considered in this work is the effect of the $\mathcal{PT}$
symmetry on PRs in the framework of the linear and nonlinear versions of
system (\ref{v2}). First, in Sections II.A-B we develop an analytical
approach for the case of moderate $\omega $, predicting at what values of $%
V_{0}$ the ac drive with fixed $\omega $ gives rise to the PR at
infinitesimally small $V_{1}$, and how tongues of the parametric instability
expand with the increase of $V_{1}$. Comprehensive dynamical charts of the
linearized version of system (\ref{v2}) and the full \ nonlinear system,
built on the basis of systematic numerical simulations, are presented in
Sections II.C-D, respectively. The structure of the charts at large $V_{1}$
is very different from the picture of PRs in usual (non-$\mathcal{PT}$)
driven dynamical systems. In Section III we present, in a brief form,
solutions of the system for slow (small $\omega $) and fast (large $%
\omega $) time-periodic modulations, which can be treated, respectively, by
means of the adiabatic and averaging approximations. The paper is concluded
by Section IV.

\section{ The analysis for moderate driving frequencies $\protect\omega $}

\subsection{Perturbative analysis of the linear system}

With a small amplitude of the ac drive, $V_{1}$, a perturbative solution to
the linearized version of Eqs. (\ref{v2}), with quasi-frequency $E$ (or
quasi-energy, in terms of quantum $\mathcal{PT}$-symmetric systems), is
looked for in the form of Eq. (\ref{quasi}), adopting the lowest-order
truncation for the Fourier expansion of the $t$-periodic amplitudes \cite{LL}%
,
\begin{equation}
\Psi _{A,B}(t)=\Psi _{A,B}^{(0)}+\Psi _{A,B}^{(+)}e^{-i\omega t}+\Psi
_{A,B}^{(-)}e^{+i\omega t},  \label{01}
\end{equation}%
with constants $\Psi _{A}^{(0,\pm )},\Psi _{B}^{(0,\pm )}$. Substituting
this ansatz into Eqs. (\ref{v2}) and keeping, in line with the form of
truncation (\ref{01}), the zeroth and first harmonics in the ensuing
expansion, gives rise to a system of linear homogeneous equations for the
vectorial set of six amplitudes,
\begin{equation}
\mathbf{\Psi }=\left\{ \Psi _{A}^{(0)},\ \Psi _{A}^{(+)},\ \Psi _{A}^{(-)},\
\Psi _{B}^{(0)},\ \Psi _{B}^{(+)},\ \Psi _{B}^{(-)}\right\} .  \label{Psi}
\end{equation}%
Nontrivial solutions exist under the condition that the determinant of the
system vanishes, which leads to the following equation for $E$:

\begin{equation}
\left\vert
\begin{array}{cccccc}
i-E & 0 & 0 & V_{0} & V_{1} & V_{1} \\
0 & i-E-\omega & 0 & V_{1} & V_{0} & 0 \\
0 & 0 & i-E+\omega & V_{1} & 0 & V_{0} \\
V_{0} & V_{1} & V_{1} & -i-E & 0 & 0 \\
V_{1} & V_{0} & 0 & 0 & -i-E-\omega & 0 \\
V_{1} & 0 & V_{0} & 0 & 0 & -i-E+\omega%
\end{array}%
\right\vert =0~.  \label{mat}
\end{equation}%
Each eigenvalue $E$ is associated with eigenvector $\mathbf{\Psi }$ of
matrix (\ref{mat}). Complex solutions of Eq. (\ref{mat}) for $E$ imply
instability of the linear system due to the spontaneous breaking of the $%
\mathcal{PT}$ symmetry, or excitation of a PR, or an interplay of both
mechanisms (as shown below, all these cases are possible). Real eigenvalues $%
E$ correspond to stable oscillatory solutions.

In the zeroth-order approximation, $V_{1}=0$, Eq. (\ref{mat}) yields an
obvious result: the spectrum remains real under the condition
\begin{equation}
V_{0}^{2}>\left( V_{0}^{2}\right) _{\min }\equiv 1  \label{1}
\end{equation}%
[more explicitly, $\left( V_{0}^{2}\right) _{\min }=\gamma ^{2}$, if scaling
$\gamma \equiv 1$ for the gain-loss coefficient is not fixed]. In this
approximation, the six eigenvalues are%
\begin{equation}
E=\pm \sqrt{V_{0}^{2}-1},E=\pm \omega \pm \sqrt{V_{0}^{2}-1},  \label{0}
\end{equation}%
where different $\pm $ signs are mutually independent.

\begin{figure}[tbp]
\begin{center}
\includegraphics[width=18cm,height=13cm]{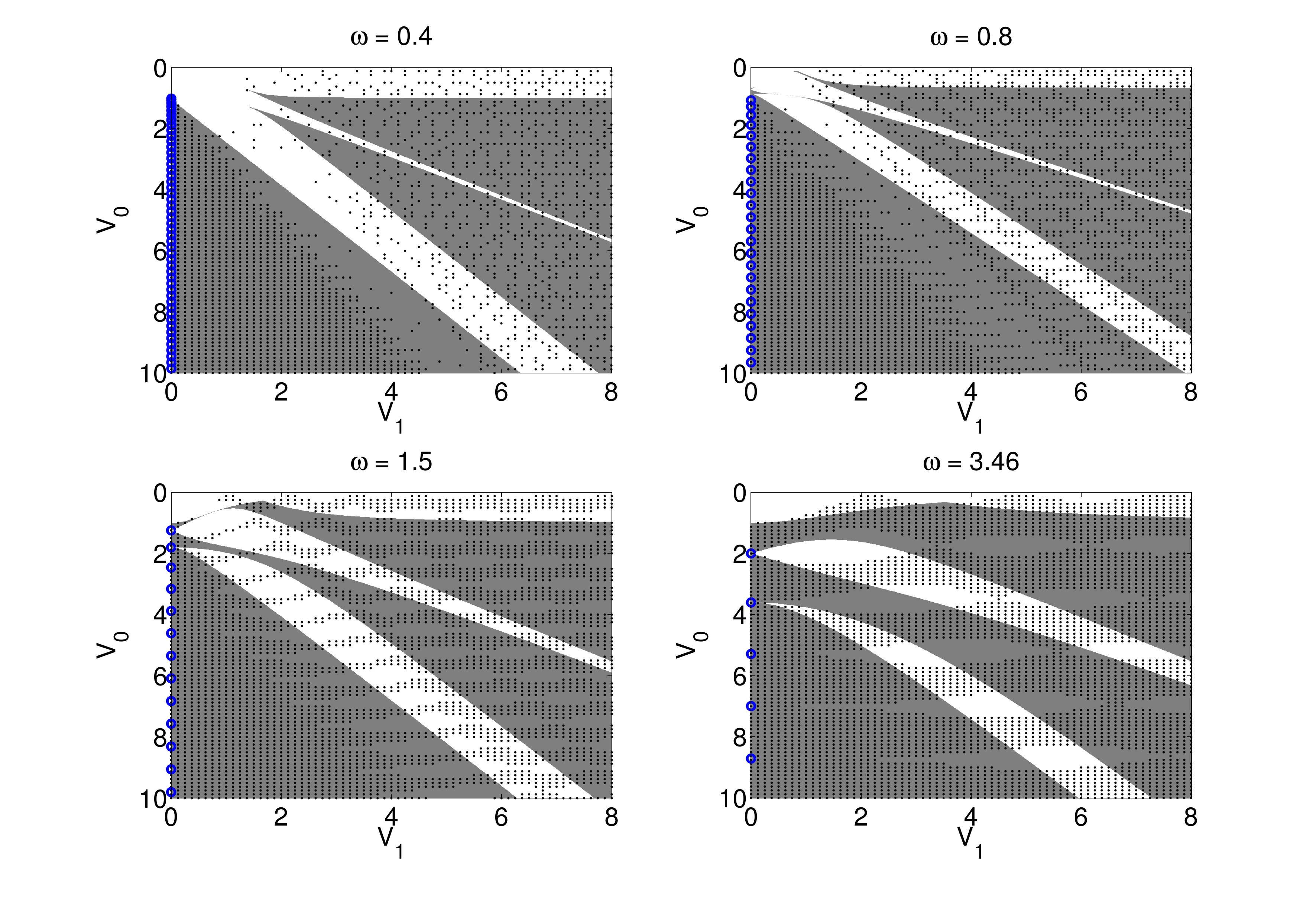} %
\end{center}
\caption{Dark and white backgrounds represent, respectively, areas of
stability and instability of the linear version of system (\protect\ref{v2})
with $\protect\chi =0$, as predicted by the numerical solution of
perturbative equation (\protect\ref{mat}), in the parameter plane of $\left(
V_{1},V_{0}\right) $ at several fixed values of $\protect\omega $ (the
equation produces solely real eigenvalues $E$ in the stable areas). The
plots are superimposed with dots denoting parameter values for which direct
simulations of the linear version of Eqs. (\protect\ref{v2}) demonstrate
stability, unstable solutions populating areas in the parameter plane which
are not covered by dots. The dotted grid was built with steps $\Delta
V_{1}=\Delta V_{0}=0.125$. Small circles on the left vertical axes,
corresponding to $V_{1}\rightarrow 0$, correspond to discrete points at
which the parametric resonances are predicted by Eq. (\protect\ref{res}).}
\label{lingrids}
\end{figure}

Stability and instability areas in the plane of the ac and dc coupling
strengths, $\left( V_{1},V_{0}\right) $, at four different values of the
driving frequency,
\begin{equation}
\omega =0.4,0.8,1.5,2\sqrt{3}\approx 3.46,  \label{omega}
\end{equation}%
as predicted by a numerical solution of algebraic equation (\ref{mat}), is
displayed in Fig. \ref{lingrids}. The plots also show stability areas
(covered by dots) as obtained in Section III by means of systematic direct
simulations of the linear version of Eqs. (\ref{v2}), i.e., the one with $%
\chi =0$. Circles on the vertical axis in Fig. \ref{lingrids} corresponds to
parametric resonances, as described in the next subsection.

\begin{figure}[tbp]
\begin{center}
\includegraphics[scale=.90]{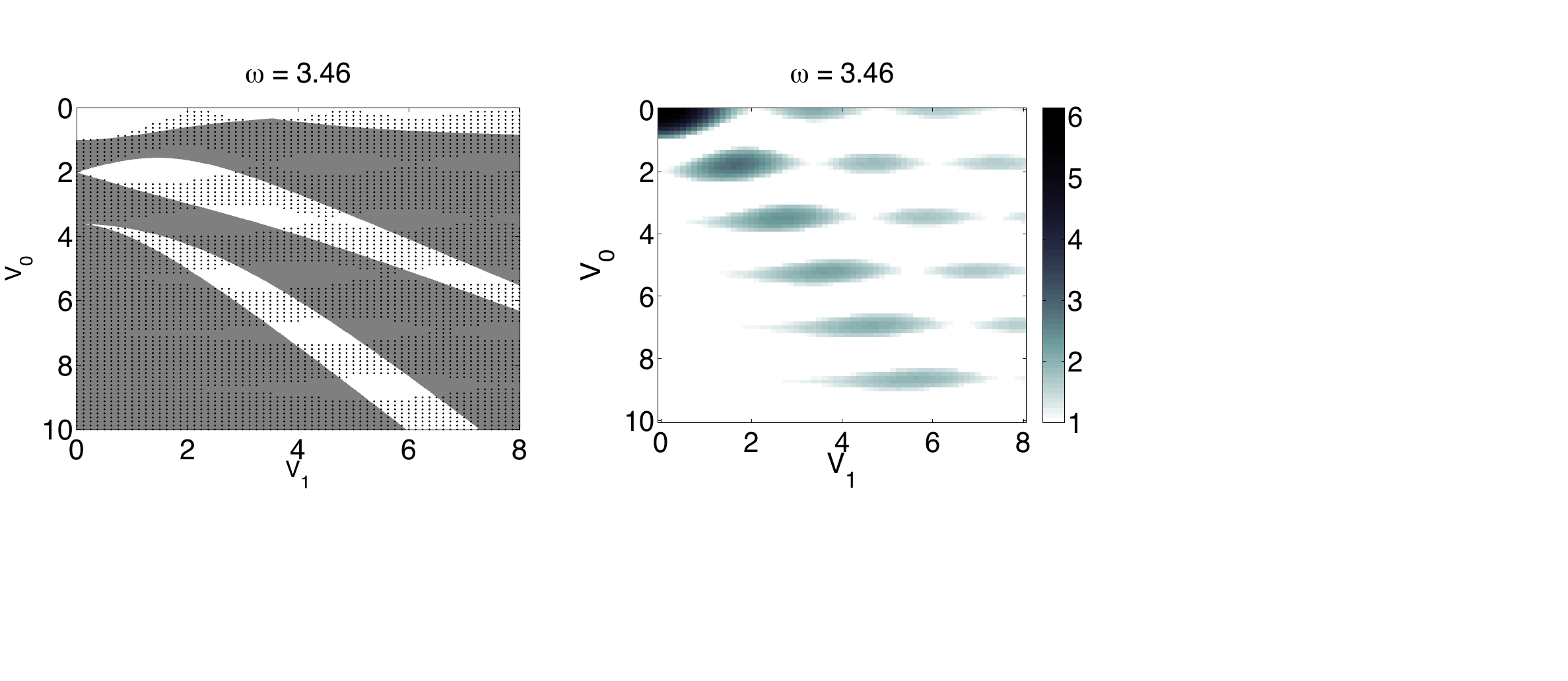}\vspace{-1in}
\end{center}
\caption{The same as the bottom right panel of Fig. \ref{lingrids} but the dots are generated by calculating the Floquet multipliers (eigenvalues) of the monodromy matrix $J_P$.  The right panel shows the magnitude of the maximal multiplier in order to demonstrate the growth rate of the instabilities. The white regions of the right panel indicate that all the Floquet multipliers are of modulus one indicating stability.
}
\label{lingrid_floq}
\end{figure}

\subsection{Parametric resonances}

In the limit of $V_{1}\rightarrow 0$, when the eigenfrequency of free
oscillations in the linearized system is $E=\sqrt{V_{0}^{2}-1}$, see Eq. (%
\ref{0}), the standard theory \cite{LL} produces the PR (i.e., the onset of
the instability of the linear system, different from the instability due to
breaking of the $\mathcal{PT}$ symmetry) at the following values of the
driving frequency:%
\begin{equation}
\omega _{\mathrm{res}}^{(n)}=\frac{2}{n}\sqrt{V_{0}^{2}-1},  \label{res}
\end{equation}%
where $n=1,2,3,...$ is the order of the resonance. Thus, for given driving
frequency $\omega $, Eq. (\ref{res}) predicts the resonance at
\begin{equation}
\left( V_{0}\right) _{\mathrm{res}}^{(n)}=\sqrt{1+\left( n\omega /2\right)
^{2}}.  \label{V0}
\end{equation}%
In Fig. \ref{lingrids}, circles on the vertical axis in each plot mark
positions of the PR as predicted by Eq. (\ref{V0}).

Note that in the case of $\omega =0.4$, Eq. (\ref{V0}) yields values
\begin{equation}
\left( V_{0}\right) _{\mathrm{res}}^{(1)}=\sqrt{1.04}\approx 1.02,~\left(
V_{0}\right) _{\mathrm{res}}^{(2)}=\sqrt{1.16}\approx 1.08,  \label{12}
\end{equation}%
which are close to the above-mentioned critical value, $\left( V_{0}\right)
_{\min }=1$, induced by the balanced combination of the gain and loss terms,
see Eq. (\ref{1}). The wide and narrow instability stripes, observed in the
panel of Fig. \ref{lingrids} corresponding to $\omega =0.4$, originate from
two critical values (\ref{12}) at small values of $V_{1}$. This situation
actually implies a \emph{merger} of the $\mathcal{PT}$-breaking and
PR-induced instabilities for relatively small values of $\omega $.

At larger $\omega $, the $\mathcal{PT}$ and PR instabilities are clearly
separated as $V_{1}\rightarrow 0$. For instance, at $\omega =2\sqrt{3}%
\approx \allowbreak 3.46$, Eq. (\ref{V0}) yields $\left( V_{0}\right) _{%
\mathrm{res}}^{(1)}=2$ for the fundamental parametric resonance ($n=1$). In
accordance with this, the respective panel of Fig. \ref{lingrids} features a
separate instability tongue, originating, at small $V_{1}$, from point $%
V_{0}=2$, and expanding with the increase of $V_{1}$. The additional tongue
observed in the same panel, originating from $V_{0}=\sqrt{13}\approx
\allowbreak 3.6$ at $V_{1}=0$, is predicted by Eq. (\ref{V0}) with $n=2$.
These tongues are quite similar to those known in the usual (non-$\mathcal{PT%
}$) model of the parametric instability.

\subsection{Direct simulations of the linear system on the parameter grid}

To produce an accurate stability chart for the linear version of the
underlying system, i.e., Eqs. (\ref{v2}) with $\chi =0$, we ran systematic
simulations of the equations on a square grid in the plane of $\left(
V_{1},V_{0}\right) $ with limits $0\leq V_{1}\leq 8$, $0\leq V_{0}\leq 10$
and steps $\Delta V_{1}=\Delta V_{0}=0.125$. We show results at the four
fixed values of the driving frequency from set (\ref{omega}). Initial values
$\left\{ \psi _{A}(0),\psi _{B}(0)\right\} $ for each run of the simulations
were calculated as per Eqs. (\ref{quasi}) and (\ref{01}), with set (\ref{Psi}%
) calculated as an eigenvector of matrix (\ref{mat}). The simulations were
performed by means of a standard fourth-order Runge-Kutta code. To check the
accuracy of the calculations, we verified that the numerical solutions would
satisfy balance equation (\ref{D}).

To distinguish
between stable and unstable solutions in the present system, we checked
whether the maximum of $|\psi _{A}(t)|$ or $|\psi _{B}(t)|$ would double
after a sufficiently long period of time, in comparison to the modulation
period, $2\pi /\omega $. It has been found that the lack of the doubling is
an appropriate indicator of the stability of a given solution (in other
words, the solutions which could be identified as stable ones would never
double their largest absolute value). Each parameter set categorized as
stable is marked by a dot in Fig. \ref{lingrids}. Unstable solutions were
identified as those breaking the above-mentioned no-doubling condition,
which was followed by visual checks, which corroborate that the so
identified unstable solutions indeed suffer blowup. In Fig. \ref{lingrids},
unstable solutions are represented by missing dots on the selected grid.

The stability of $t$-periodic solutions can also be numerically
analyzed by computing the eigenvalues of the monodromy matrix $J_P$
for small perturbations around the given solution \cite{perturbation}
(in the case of the linear problem, this effectively amounts to the
stability
of the zero solution).   Writing (\ref{v2}) with $\chi = 0$ as a system of four real equations with Jacobian $J_F(t)$ implies that $J_P$, the Jacobian of the Poincar{\'e} map, satisfies the variational equation
\begin{equation}
\label{JPvar}
\frac{d}{dt} (J_P) = J_F (J_P)
\end{equation}
with initial condition $J_P(0) = Id$.  Solving (\ref{JPvar}) and
computing the eigenvalues of $J_P(t=2\pi/\omega)$ gives the Floquet
multipliers which indicate instability if any of the multipliers have
modulus greater than one.  In Fig. \ref{lingrid_floq} we show that
this Floquet multiplier 
calculation predicts the {\it exact} same (in)stability intervals as
that which we observed using the doubling condition of our direct
integration as explained above.

Figure \ref{lingrids} clearly demonstrates both similarities and nontrivial
differences between the present stability charts and those well known in the
PR theory. Indeed, different numerical stability tongues originating at $%
V_{1}=0$ do not bend down, as in the usual theory, but keep the direction
parallel to the axis of $V_{1}$. This difference suggests that the
truncation adopted in Eq. (\ref{01}) becomes irrelevant at large $V_{1}$, as
a larger number of harmonics get involved into the dynamics, as corroborated
by spectra of particular solutions displayed below. With the decrease of $%
\omega $, the number of tongues becomes large, while their widths and gaps
between adjacent ones get smaller. At small $\omega $, the stability chart
features a complex small-scale structure,
which has no counterpart in the ordinary PR theory. It is important to point
out that, while the quasi-energy-based analysis presented in Section II.A is
expected (and indeed found in Fig.~\ref{lingrids}) to be valid for small $%
V_{1}$, the PR theory of section II.B is relevant in a different limit. In
particular, it accurately captures the dynamics at large $\omega $, where
the PR points are no longer clustered, and their tongues shrink accordingly.
It is clearly observed in Fig. \ref{lingrids} that the horizontal
tongues for the cases of, e.g., $\omega =1.5$ and $3.46$ naturally emanate
from the PR points of $V_{1}\rightarrow 0$. An additional relevant point to
mention in that light is that higher-order resonances appear to be initiated
\textquotedblleft deeper\textquotedblright\ along the $V_{1}$ axis, although
this observation may be affected by a finite extension of our simulations,
and of a weak growth rate of these instabilities at small $V_{1}$.

\begin{figure}[tbp]
\begin{center}
\includegraphics[width=18.5cm,height=11cm]{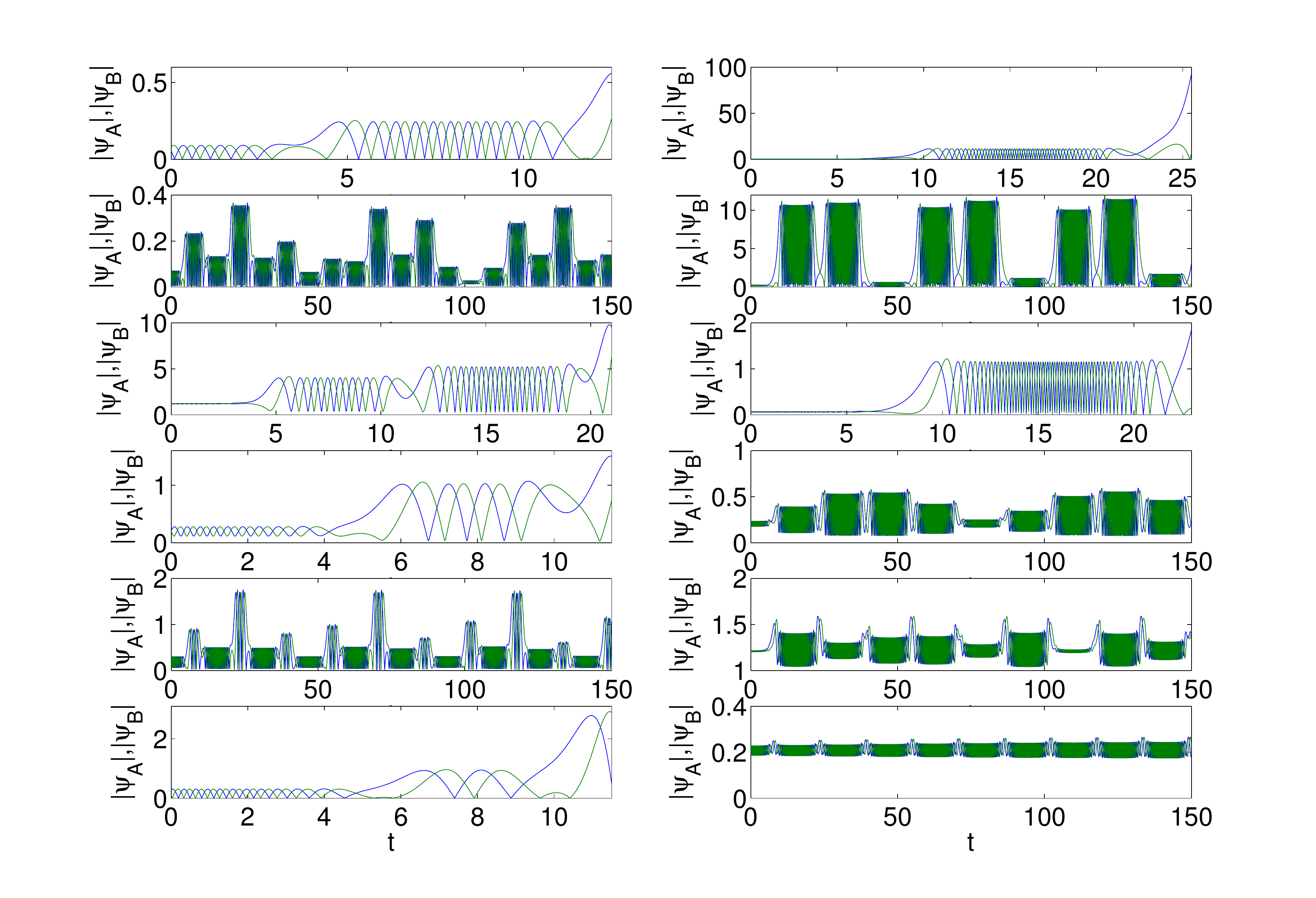} %
\end{center}
\caption{Solutions $|\protect\psi _{A}(t)|,|\protect\psi _{B}(t)|$ produced
by the numerical simulations of the linear version of Eqs.~(\protect\ref{v2}%
) for parameter values corresponding to a vertical cut in Fig. \protect\ref%
{lingrids} at $\omega = 0.4$ and $V_{1}=3.25$. These plots confirm that actually present black
dots and missing ones (empty sites of the grid) in Fig. \protect\ref%
{lingrids} represent stable and unstable solutions, respectively. Values of $%
V_{0}$ corresponding to the left and right columns (from top to bottom) are $%
0,0.5,1,3,3.5$, $4$, and $5.5,6,6.5,7.5,8,8.5$, respectively.}
\label{vertcutprofiles}
\end{figure}

Figure \ref{vertcutprofiles} shows characteristic examples of the solutions
for select combinations of parameter values, both stable (quasi-periodic
solutions) and unstable (growing ones). Due to the presence of the gain in
the system, unstable solutions grow indefinitely, with (at most) an
exponential rate~\cite{pelinov}.
As a result, at some point in time, after the temporal interval displayed in
respective panels of Fig. \ref{vertcutprofiles}, each amplitude increases
rapidly (eventually the evolution accrues numerical errors).

\begin{figure}[tbp]
\begin{center}
\includegraphics[scale=0.45]{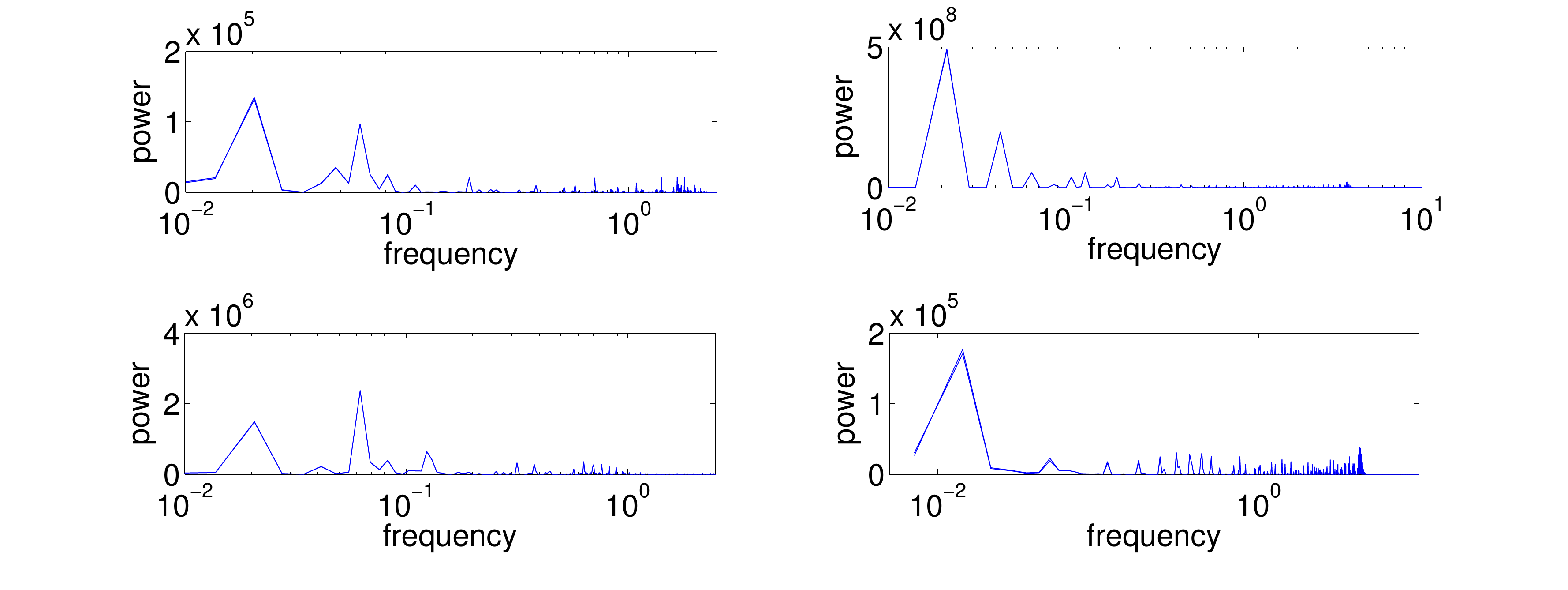}
\end{center}
\caption{For $\protect\omega =0.4$, plotted on the logarithmic scale of the
frequency are power spectra computed as per Eq. (\protect\ref{FFT}) for the
following values of the parameters: $(V_{1},V_{0})=$ $(3.25,0.5)$ in the top
left, $(3.25,3.5)$ in the bottom left, $(3.25,6)$ in the top right, and $%
(3.25,7.5)$ in the bottom right. The frequency in the plots is $f$ defined
in Eq. (\protect\ref{FFT}), and the power spectra for both components of
stable solutions completely overlap. The corresponding solutions $\protect%
\psi _{A,B}(t)$ are displayed in Fig. \protect\ref{vertcutprofiles}. A
prominent peak is found at frequency $0.0637=(\protect\omega =0.4)/2\protect%
\pi $.}
\label{vertcutfft}
\end{figure}

For a few stable oscillatory solutions, Fig. \ref{vertcutfft} shows the
power spectra generated by means of the DFFT (discrete fast Fourier
transform),%
\begin{equation}
\mathrm{power~}_{A,B}(f_{k})=\left\vert \frac{1}{N}\sum_{j=1}^{N}\psi
_{A,B}(t_{j})\exp \left( 2\pi i\frac{(j-1)k}{N}\right) \right\vert ^{2},
\label{FFT}
\end{equation}%
where $t_{j}=j\Delta t$, $t_{N}$ is the total integration time, and $%
f_{k}\in \frac{1}{N\Delta t}\left\{ 1,\dots ,N/2\right\} $ for even $N$. All
the spectra show a prominent peak at the driving frequency, $\omega /\left(
2\pi \right) $. One also observes a set of weaker higher-frequency peaks
corresponding to the frequency multiplication by the parametric drive.

To resolve the higher frequencies (smaller periods), we have performed the
FFT on smaller windows. For example, within the first few seconds of the
simulations for small $\omega $, we would observe a dominant frequency in
the spectrum consistent with the prediction, $f=\sqrt{(V_{0}+2V_{1})^{2}-1}%
/(2\pi )$, given by Eq. (\ref{0}) with $\omega =0$. As $\psi _{A}(t)$ and $%
\psi _{B}(t)$ evolve in time, the high frequencies lead to modulations. For
parameter values $\left( V_{0},V_{1}\right) $ which are buried deepest in
stability regions in Fig. \ref{lingrids}, the solutions feature the least
amount of such fluctuations at high frequencies. These fluctuations account
for the small-amplitude pattern in the high-frequency part of the spectra in
Fig. \ref{vertcutfft}, which appears when the FFT is performed on a large
interval, such as $[0,250]$. All of the frequencies which appear in Fig. \ref%
{vertcutfft} are linear combinations of those predicted at $\omega =0$ and
the driving frequency, $\omega /\left( 2\pi \right) $.

\subsection{The nonlinear propagation}

As indicated above, the initial values of $\left( \psi _{A},\psi _{B}\right)
$, used in the simulations of the linear version of Eqs. (\ref{v2}), were
generated as per Eqs. (\ref{quasi}) and (\ref{01}), into which eigenvectors (%
\ref{Psi}) of the matrix from Eq. (\ref{mat}) were substituted. We now use
the same inputs for the systematic simulations of the full nonlinear form of
Eqs. (\ref{v2}) with $\chi =1$.

\begin{figure}[tbp]
\begin{center}
\includegraphics[width=18cm,height=13cm]{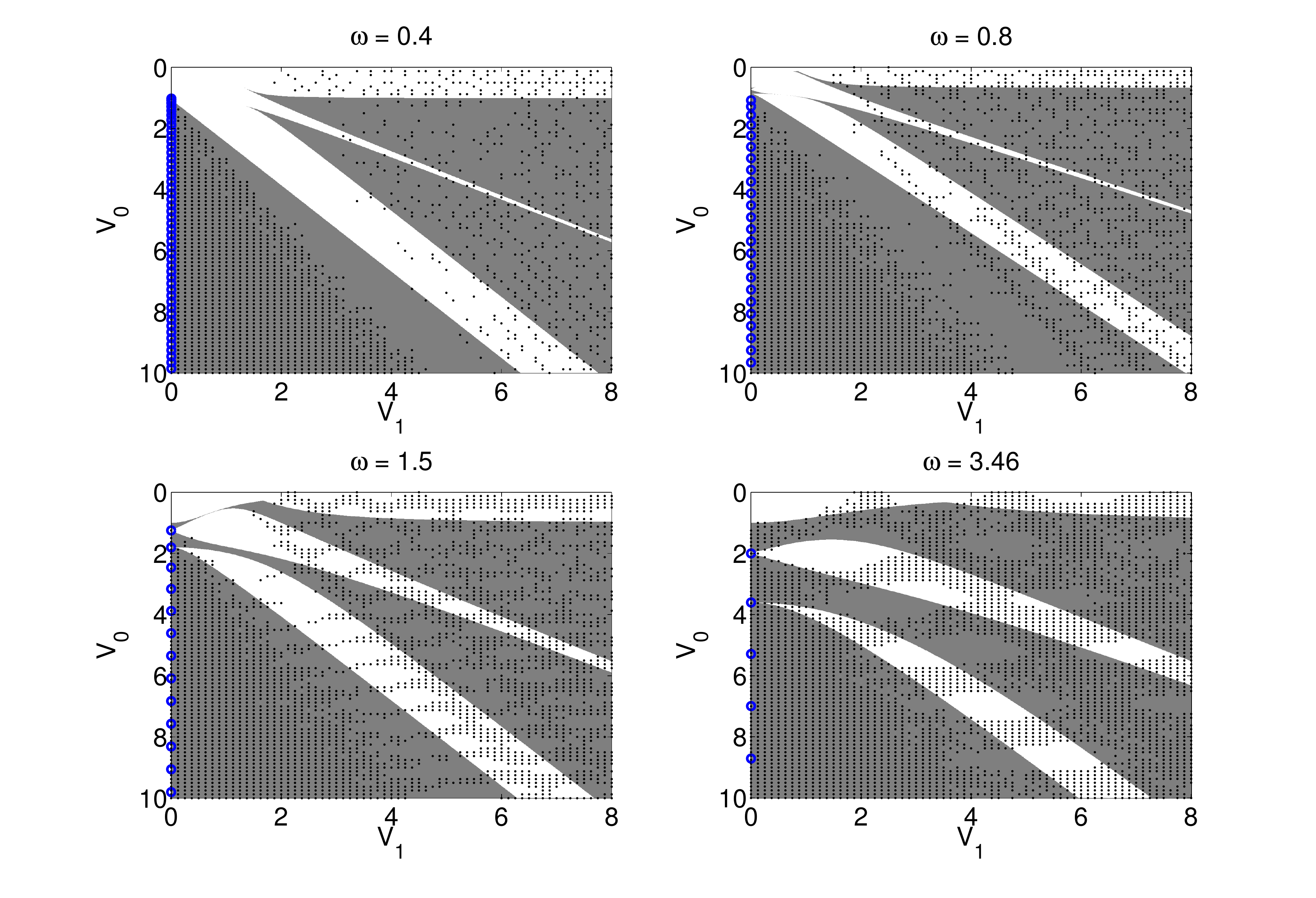} %
\end{center}
\caption{The same as in Fig. \protect\ref{lingrids}, but with dots
representing robust numerical solutions of the full nonlinear equations (%
\protect\ref{v2}). See Figs. \protect\ref{nlinvertcut1} and \protect\ref%
{nlinvertcut2} for profiles of typical solutions. }
\label{nlingrids}
\end{figure}

Figure \ref{nlingrids} shows the so-generated nonlinear counterpart of Fig. %
\ref{lingrids}, on the same grid of parameter values. In the upper two
predicted-as-stable (gray) regions of Fig. \ref{nlingrids}, the tests
generally show that the solutions are more likely to be unstable, in
comparison with the linear system. In the inner portions of the lower gray
regions, many solutions stay stable in the presence of the nonlinearity,
while they are more likely to be unstable at the boundary of the region, in
comparison with the linear case.

\begin{figure}[tbp]
\begin{center}
\includegraphics[width=18.5cm,height=11cm]{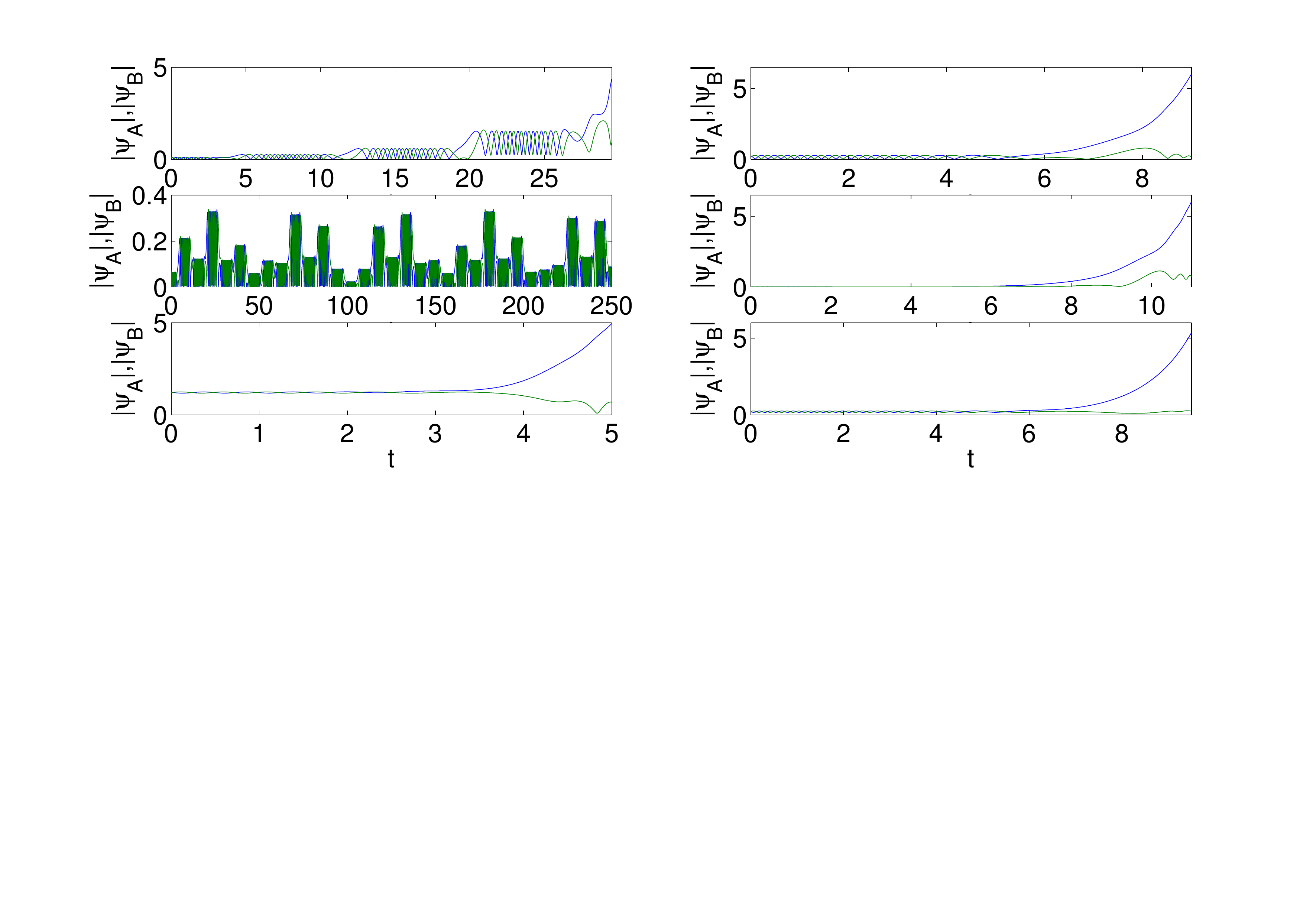} %
\vspace{-2.25in}
\end{center}
\caption{ Each plot shows solutions evolving according to nonlinear
equations (\protect\ref{v2}). Parameter values $(V_{1},V_{0})$ follow the
same pattern as the top six plots in Fig. \protect\ref{vertcutprofiles}: $%
(3.25,0),(3.25,0.5),(3.25,1)$ on the left, and $%
(3.25,5.5),(3.25,6),(3.25,6.5)$ on the right. The respective initial
amplitudes of each solution are: $(|\protect\psi _{A}|,|\protect\psi %
_{B}|)=(0.0530,0.0868)$, $(0.0225,0.0712)$, $(1.2052,1.2052)$, $%
(0.2516,0.1538)$, $(0.2087,0.2087)$, $(0.0607,0.0607).$ Of the two stable
solutions of the linear system, only one remains stable in the nonlinear
case. }
\label{nlinvertcut1}
\end{figure}

Some stable solutions of the linear system remain apparently robust when the
nonlinearity is added, but with different amplitudes, in comparison to their
linear counterparts. Profiles shown in Figs. \ref{nlinvertcut1} and \ref%
{nlinvertcut2} are nonlinear extensions of those displayed above in Fig. \ref%
{vertcutprofiles}. As in the case of the linear system, the balance equation
(\ref{D}) remains valid for stable solutions, while the unstable solutions
grow indefinitely with the (typically) exponential growth rate predicted in~%
\cite{pelinov}. 

\begin{figure}[tbp]
\begin{center}
\includegraphics[width=18.5cm,height=11cm]{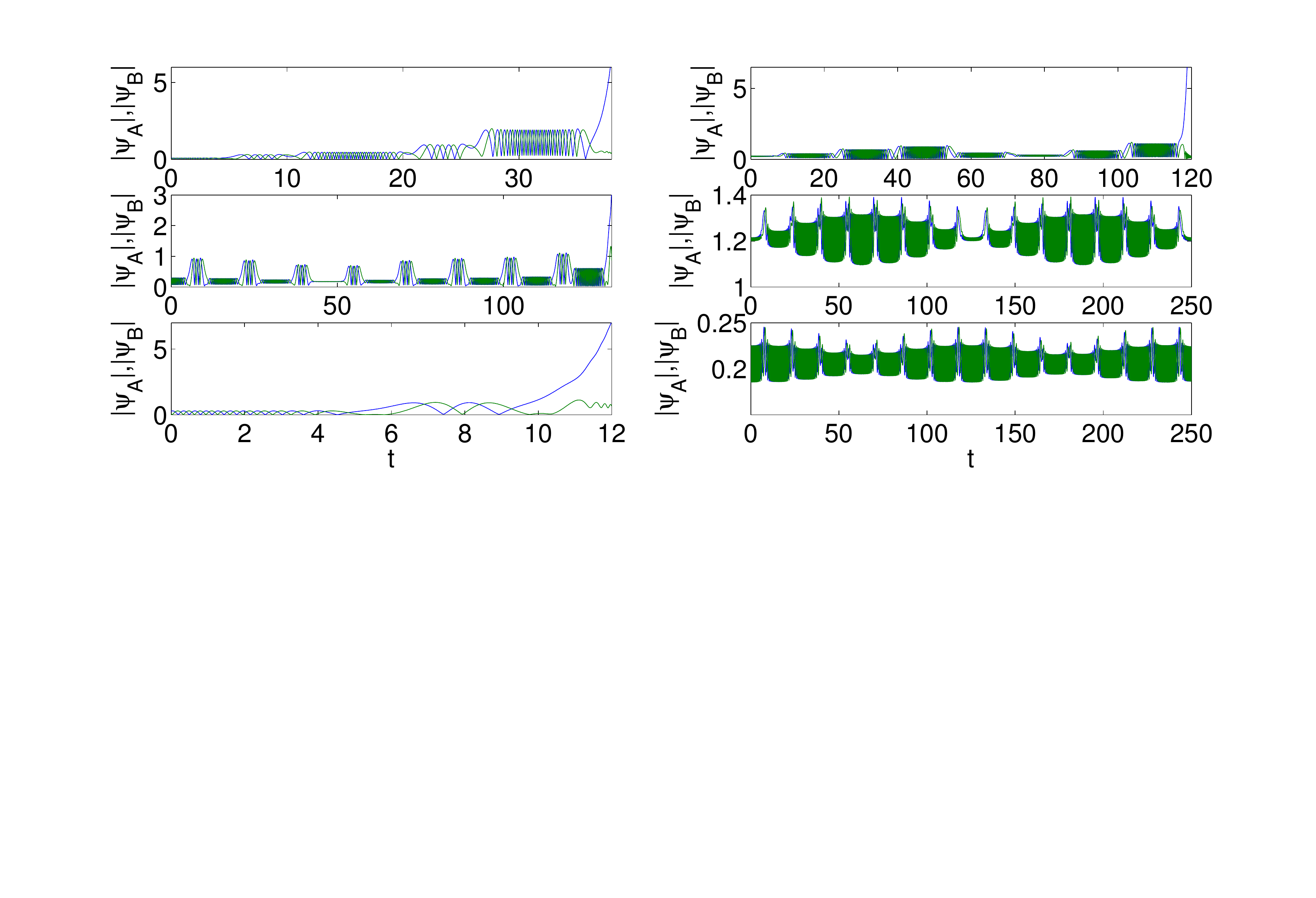} %
\vspace{-2.25in}
\end{center}
\caption{The same as in Fig. \protect\ref{nlinvertcut1}. Values $%
(V_{1},V_{0})$ follow the same pattern as the bottom six plots of Figure
\protect\ref{vertcutprofiles}: $(3.25,3)$, $(3.25,3.5)$, $(3.25,4)$, $%
(3.25,7.5)$, $(3.25,8)$, $(3.25,8.5)$. Respective initial amplitudes of the
wave functions are $(|\protect\psi _{A}|,|\protect\psi %
_{B}|)=(0.2095,0.2095) $, $(0.2098,0.2098)$, $(0.2098,0.2098)$, $%
(0.2089,0.2089)$, $(0.2089,0.2089)$, $(1.2052,1.2052)$. Out of the four
stable solutions of the corresponding linear system, two become unstable,
and two remain stable, with amplitudes differing from their counterparts in
the linear system.}
\label{nlinvertcut2}
\end{figure}

Finally, it is relevant to stress that the addition of the nonlinearity does
not necessarily lead to a partial destabilization of the solutions, as the
effect may be the opposite (a possibility of a partial stabilization of the $%
\mathcal{PT}$ symmetry under the action of nonlinearity was also recently
demonstrated, in a different context, in Ref. \cite{Moti}. In particular,
there are a few dots in Fig. \ref{nlingrids} that do not appear in Fig. \ref%
{lingrids}. Profiles of a few solutions in the linear and nonlinear systems,
corresponding to three of these dots, are displayed in Fig. \ref%
{nlinstabilizes}.

\begin{figure}[tbp]
\begin{center}
\includegraphics[width=18.5cm,height=11cm]{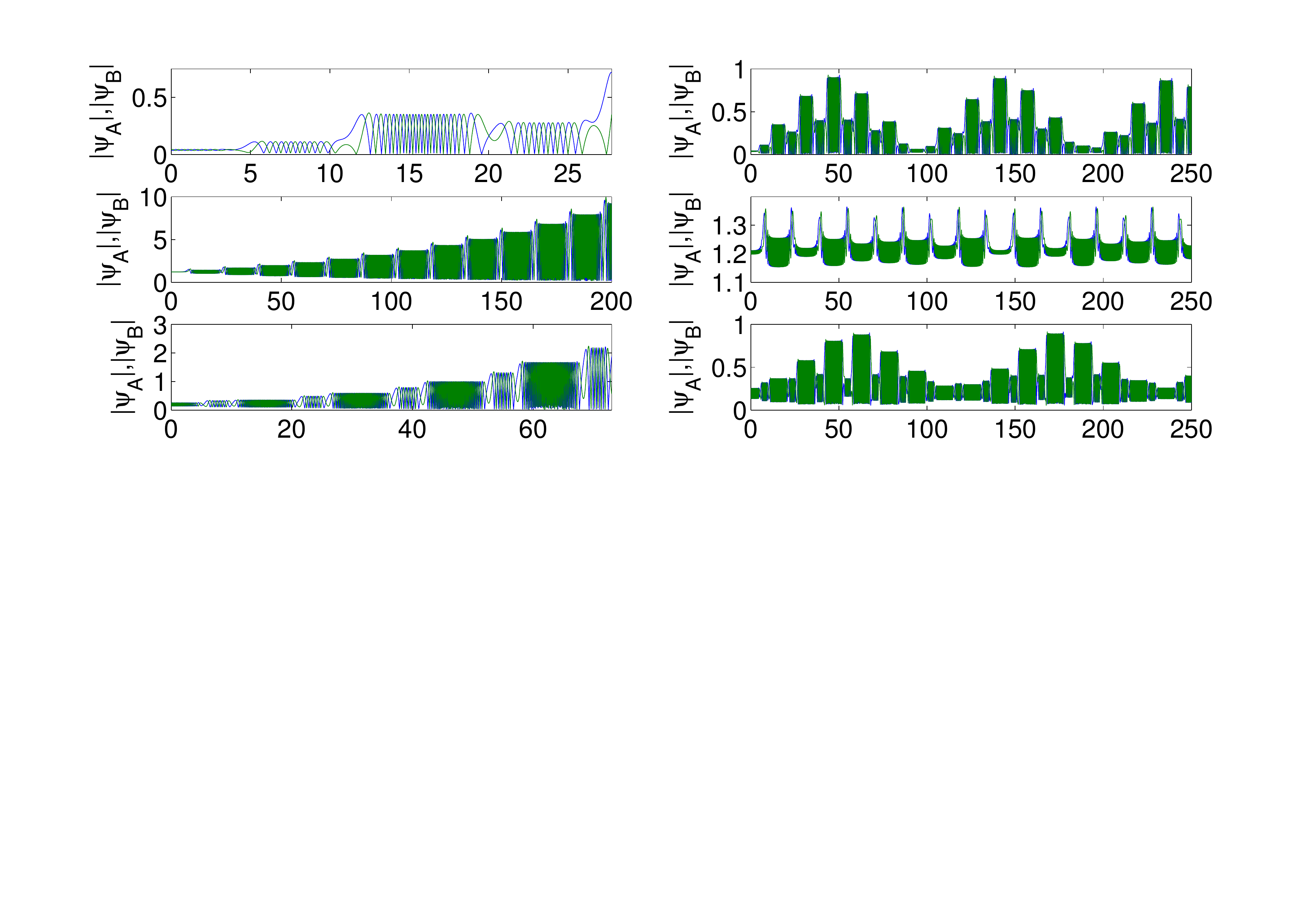} %
\vspace{-2.25in}
\end{center}
\caption{ Plots of $|\protect\psi _{A}(t)|,|\protect\psi _{B}(t)|$ are shown
at parameter values which correspond to missing dots in Fig. \protect\ref%
{lingrids} (unstable in the linear system) and dots in Fig. \protect\ref%
{nlingrids} (stable in the nonlinear system) for $\omega=0.4$. The left and right columns
correspond, respectively, to the linear and nonlinear propagation. Values $%
(V_{1},V_{0})$ from top to bottom are $(3.5,1.5)$, $(4,9.5)$, $(7,8)$. For
the nonlinear simulations, initial values are $(|\protect\psi _{A}|,|\protect%
\psi _{B}|)=(0.0437,0.0437)$, $(1.2059,1.2059)$, $(0.2076,0.2076)$. }
\label{nlinstabilizes}
\end{figure}


\section{The cases of slow and fast modulations\newline (adiabatic and averaging
approximations) }

In addition to the analysis presented above, some simple results can be
obtained, like in other dynamical models \cite{Itin}, by means of the
adiabatic approximation, which makes it possible to study some specific
instabilities of the corresponding nonlinear systems \cite{Itin2}.
{ In both the adiabatic approximation and the averaging method for
fast modulations given below, we will attempt to reveal
the dependence of the solution on
relevant parameters such as $V_0$, $V_1$, $\chi$ and $\omega$.}

The
adiabatic approximation is valid for very small frequencies $\omega $, when
eigenvalues (\ref{0}) are nearly degenerate, and
\begin{equation}
\tilde{V}_{0}(t)\equiv V_{0}+2V_{1}\cos \left( \omega t\right)
\label{tilde}
\end{equation}%
is a slowly varying function of $t$. Treating coefficient (\ref{tilde}) as a
constant, solutions of nonlinear system (\ref{v2}) with $\chi =1$ then
follow from known results \cite{KPZ}: setting
\begin{equation}
\psi _{A}=e^{-iEt}C,\psi _{B}=e^{-iEt}C^{\ast },  \label{solution}
\end{equation}%
one finds two solutions,
\begin{eqnarray}
C_{1} &=&\sqrt{E-\sqrt{\tilde{V}_{0}^{2}-1}}\exp \left[ \frac{i}{2}\arcsin
\left( \frac{1}{\tilde{V}_{0}}\right) \right] ,  \label{C1} \\
C_{2} &=&\sqrt{E+\sqrt{\tilde{V}_{0}^{2}-1}}i\exp \left[ -\frac{i}{2}\arcsin
\left( \frac{1}{\tilde{V}_{0}}\right) \right] .  \label{C2}
\end{eqnarray}

These solutions are valid as long as conditions $\tilde{V}_{0}^{2}(t)>1$ and
$E>\pm \sqrt{\tilde{V}_{0}^{2}(t)-1}$ hold at all $t$, with the upper and
lower signs corresponding to solutions (\ref{C1}) and (\ref{C2}),
respectively. Combining these conditions for $C=C_{1}$ solution produces a
validity range,
\begin{equation}
1+2V_{1}<V_{0}<\sqrt{1+E^{2}}-2V_{1},  \label{-range}
\end{equation}%
which exists as long as $E>E_{\min }\equiv 4V_{1}\left( 1+V_{1}\right) $.
For the $C=C_{2}$ solution, the above conditions amount to
\begin{equation}
V_{0}>\left\{
\begin{array}{c}
2V_{1}+1,~~\mathrm{for}~~E>0, \\
2V_{1}+\sqrt{1+E^{2}},~~\mathrm{for}~~E<0.%
\end{array}%
\right.   \label{+range}
\end{equation}

The above solutions correspond to those reported in Ref. \cite{KPZ} by
setting $\gamma =1$ and $k=\tilde{V_{0}}$. In the limit of $\omega =0$,
solution (\ref{C2}) is known to be stable, which is corroborated by our
simulations of the system at small $\omega \neq 0$, with parameters from
range (\ref{+range}). On the other hand, it is known too \cite{KPZ} that, at
$\omega =0$, solution (\ref{C1}) is stable only for $\tilde{V_{0}}$
satisfying an additional condition,
\begin{equation}
\tilde{V}_{0}\geq \sqrt{1+E^{2}/4}.  \label{extra}
\end{equation}

In direct simulations performed at $\omega \neq 0$, the actually observed
stability boundary rises to values exceeding the one (\ref{extra}). For
example, if $V_{1}=1/2$ and $E=\sqrt{15}$, the lower existence bound for
solution (\ref{C1}) in Eq. (\ref{-range}) is $V_{0}>1+2V_{1}=2$, whereas Eq.
(\ref{extra}) yields $V_{0}>\sqrt{1+E^{2}/4}-2V_{1}\approx 1.179$, which is
immaterial, as the existence threshold is higher. Nevertheless, even for $%
\omega $ as small as $0.001$ direct simulations reveal a higher instability
boundary, at $V_{0}\approx 2.4$. Only at extremely small values of $\omega $
the numerically found the instability boundary drops down to its value (\ref%
{extra}) predicted for the static system.

The opposite case of fast modulations in system (\ref{v2}) implies the
consideration of large $\omega $, when the averaging method may be applied
\cite{averaging}. The corresponding solution is easily obtained in the first
approximation with respect to small $1/\omega $:%
\begin{equation}
\left\{
\begin{array}{c}
\psi _{A} \\
\psi _{B}%
\end{array}%
\right\} =\left\{
\begin{array}{c}
\psi _{A}^{(0)} \\
\psi _{B}^{(0)}%
\end{array}%
\right\} -\frac{2iV_{1}}{\omega }\sin \left( \omega t\right) \left\{
\begin{array}{c}
\psi _{B}^{(0)} \\
\psi _{A}^{(0)}%
\end{array}%
\right\} ,  \label{1/omega}
\end{equation}%
where the unperturbed solution is given by Eqs. (\ref{solution}) and (\ref%
{C1}), (\ref{C2}) with $V_{1}=0$ (provided that condition $V_{0}^{2}>1$
holds). Thus, the respective final results are
\begin{equation}
\left\{
\begin{array}{c}
\psi _{A}^{(0)} \\
\psi _{B}^{(0)}%
\end{array}%
\right\} =\pm e^{-iE t}\sqrt{E -\sqrt{V_{0}^{2}-1}}\left\{
\begin{array}{c}
\exp \left[ \frac{i}{2}\arcsin \left( \frac{1}{V_{0}}\right) \right]  \\
\exp \left[ -\frac{i}{2}\arcsin \left( \frac{1}{V_{0}}\right) \right]
\end{array}%
\right\} ;  \label{V1=0+}
\end{equation}%
\begin{equation}
\left\{
\begin{array}{c}
\psi _{A}^{(0)} \\
\psi _{B}^{(0)}%
\end{array}%
\right\} =\pm ie^{-iE t}\sqrt{E +\sqrt{{V}_{0}^{2}-1}}%
\left\{
\begin{array}{c}
\exp \left[ \frac{i}{2}\arcsin \left( \frac{1}{V_{0}}\right) \right]  \\
-\exp \left[ -\frac{i}{2}\arcsin \left( \frac{1}{V_{0}}\right) \right]
\end{array}%
\right\} .  \label{V1=0-}
\end{equation}

\section{Conclusions}

We have introduced the simplest model which opens the way to studying the
concept of quasi-energy and PRs (parametric resonances), with ensuing
instability, in $\mathcal{PT}$-symmetric systems, that incorporate spatially
separated and mutually balanced gain and loss elements. The parametric drive
is introduced in the form of the periodic (ac) modulation of the coefficient
of the coupling between the two degrees of freedom. {The main results are
reported for moderate values of driving frequency $\omega $. T}he onset and
development of the parametric instability at an infinitesimal amplitude of
the ac drive was predicted analytically. The full picture was obtained by
means of systematic simulations of both linear and nonlinear versions of the
system. At large values of $\omega $, the tongues of the parametric
instability originate, with the increase of the driving amplitude ($V_{1}$),
at values of the coupling constant ($V_{0}$) predicted by the analytical
consideration of the resonance points. However, the continuation of the PR
tongues with the increase of $V_{1}$ in the $\mathcal{PT}$-symmetric system
follows a scenario which is very different from the usual parametrically
driven system: instead of bending down to larger values of $V_{0}$, the
tongues continue in directions parallel to the $V_{1}$ axis. The decrease of
$\omega $ makes the system of parallel tongues denser, eventually replacing
the relatively simple system of parallel instability tongues by a complex
small-scape structure. While the parametric resonance picture is an accurate
predictor of the tongue formation at large $\omega $, the quasi-energy
method yields reasonably accurate results at small $V_{1}$. The inclusion of
the nonlinearity gradually deforms the picture, destabilizing a part of
originally stable dynamical regimes, and stabilizing a few originally
unstable ones. Nevertheless, as it might be expected, the chief effect of
the nonlinearity is the expansion of the instability area. We have also
presented a short account solutions obtained by means of the adiabatic and
averaging approximations, for very slow and fast modulations, respectively.

This analysis may be extended in various ways, adding more terms to the
system. In particular, a straightforward generalization may include
nonlinear gain and loss terms, in addition to the linear pair of such terms,
cf. Refs. \cite{AKKZ,Canberra,rudy}. Extensions beyond the dimer case to
oligomers and eventually to lattices would also be particularly interesting
to consider~\cite{discrete,circular,KPZ,pelinov}. In particular, discrete $%
\mathcal{PT}$-symmetric solitons may be expected in the latter case.
These themes are currently under consideration and will be presented in
future publications.

\section*{Acknowledgment}

This work was supported, in a part, by the Binational (US-Israel) Science
Foundation through grant No. 2010239.  PGK gratefully also acknowledges support from NSF-DMS-1312856, NSF-CMMI-1000337
and AFOSR-FA9550-12-1-0332.


\begin{thebibliography}{99}
\bibitem{Bender_review} C. M. Bender, Rep. Prog. Phys. \textbf{70}, 947
(2007).

\bibitem{Muga} A. Ruschhaupt, F. Delgado, and J. G. Muga, J. Phys. A: Math.
Gen. \textbf{38}, L171 (2005). 

\bibitem{experiment} A. Guo, G. J. Salamo, D. Duchesne, R. Morandotti, M.
Volatier-Ravat, V. Aimez, G. A. Siviloglou, and D. N. Christodoulides, Phys.
Rev. Lett. \textbf{103}, 093902 (2009); C. E. R\"{u}ter, K. G. Makris, R.
El-Ganainy, D. N. Christodoulides, M. Segev, and D. Kip, Nature Phys.
\textbf{6}, 192 (2010).

\bibitem{Kottos} N. Bender, S. Factor, J. D. Bodyfelt, H. Ramezani, D. N.
Christodoulides, F. M. Ellis, and T. Kottos, Phys. Rev. Lett. \textbf{110},
234101 (2013); see also J. Schindler, A. Li, M. C. Zheng, F. M. Ellis and T.
Kottos, Phys. Rev. A \textbf{84}, 040101 (2011).

\bibitem{BEC} E. M. Graefe, U. G\"{u}nther, H. J. Korsch, and A. E.
Niederle, J. Phys. A:\ Math. Theor. \textbf{41}, 255206 (2008); H. Cartarius
and G. Wunner, Phys. Rev. A \textbf{86}, 013612 (2012).

\bibitem{special-issues} See special issues: H. Geyer, D. Heiss, and M.
Znojil, Eds., J. Phys. A: Math. Gen. \textbf{39}, \textit{Special Issue
Dedicated to the Physics of Non-Hermitian Operators} (\textit{PHHQP IV})
(University of Stellenbosch, South Africa, 2005) (2006); A. Fring, H. Jones,
and M. Znojil, Eds., J. Math. Phys. A: Math Theor. \textbf{41}, \textit{%
Papers Dedicated to the Subject of the 6th International Workshop on
Pseudo-Hermitian Hamiltonians in Quantum Physics} (\textit{PHHQPVI}) (City
University London, UK, 2007) (2008); C. Bender, A. Fring, U. G\"{u}nther,
and H. Jones, Eds., \textit{Special Issue: Quantum Physics with
non-Hermitian Operators}, J. Math. Phys. A: Math Theor. \textbf{41}, No. 44
(2012).

\bibitem{review} K. G. Makris, R. El-Ganainy, D. N. Christodoulides, and Z.
H. Musslimani, Int. J. Theor. Phys. \textbf{50}, 1019 (2011).

\bibitem{Musslimani2008} Z. H. Musslimani, K. G. Makris, R. El-Ganainy, and
D. N. Christodoulides, Phys. Rev. Lett. \textbf{100}, 030402 (2008); Z. Lin,
H. Ramezani, T. Eichelkraut, T. Kottos, H. Cao, and D. N. Christodoulides,
Phys. Rev. Lett. \textbf{106}, 213901 (2011); X. Zhu, H. Wang, L.-X. Zheng,
H. Li, and Y.-J. He, Opt. Lett. \textbf{36}, 2680 (2011); C. Li, H. Liu, and
L. Dong, Opt. Exp. \textbf{20}, 16823 (2012); C. M. Huang, C. Y. Li, and L.
W. Dong, \textit{ibid}. \textbf{21}, 3917 (2013).

\bibitem{Yang} S. Nixon, L. Ge, and J. Yang, Phys. Rev. A \textbf{85},
023822 (2012).



\bibitem{dark} H. G. Li, Z. W. Shi, X. J. Jiang, and X. Zhu, Opt. Lett.
\textbf{36}, 3290 (2011); V. Achilleos, P. G. Kevrekidis, D. J.
Frantzeskakis, and R. Carretero-Gonz\'{a}lez, Phys. Rev. A \textbf{86},
013808 (2012); see also V. Achilleos, P. G. Kevrekidis, D. J. Frantzeskakis,
R. Carretero-Gonz{\'{a}}lez, {\textit{ Localized Excitations in Nonlinear Complex Systems}}, R. Carretero-Gonz‡lez, J. Cuevas-Maraver, D.ÊFrantzeskakis, N.ÊKarachalios, P.ÊKevrekidis, F.ÊPalmero-Acebedo, (Eds.)
Springer-Verlag (Heidelberg, 2013).

\bibitem{chi2} F. C. Moreira, F. K. Abdullaev, V. V. Konotop, and A. V.
Yulin, Phys. Rev. A \textbf{86}, 053815 (2012); F. C. Moreira, V. V.
Konotop, and B. A. Malomed, \textit{ibid}. \textbf{87}, 013832 (2013).

\bibitem{dual} R. Driben and B. A. Malomed, Opt. Lett. \textbf{36}, 4323
(2011); N. V. Alexeeva, I. V. Barashenkov, A. A. Sukhorukov, and Y. S.
Kivshar, Phys. Rev. A \textbf{85}, 063837 (2012); I. V. Barashenkov, S. V.
Suchkov, A. A. Sukhorukov, S. V. Dmitriev, and Y. S. Kivshar, \textit{ibid}.
\textbf{86}, 053809 (2012).

\bibitem{dual2} F. K. Abdullaev, V. V. Konotop, M. \"{O}gren, and M. P. S\o %
rensen, Opt. Lett. \textbf{36}, 4566 (2011); R. Driben and B. A. Malomed,
EPL \textbf{96}, 51001 (2011); \textit{ibid}. \textbf{99}, 54001 (2012).

\bibitem{dark-coupler} Yu. V. Bludov, V. V. Konotop, and B. A. Malomed,
Phys. Rev. A \textbf{87}, 013816 (2013).

\bibitem{discrete} S. V. Dmitriev, A. A. Sukhorukov, and Y. S. Kivshar, Opt.
Lett. \textbf{35}, 2976 (2010); S. V. Suchkov, B. A. Malomed, S. V.
Dmitriev, and Y. S. Kivshar, Phys. Rev. E \textbf{84}, 046609 (2011); S. V.
Suchkov, A. A. Sukhorukov, S. V. Dmitriev, and Y. S. Kivshar, EPL \textbf{100%
}, 54003 (2012); D. A. Zezyulin and V. V. Konotop, Phys. Rev. Lett. \textbf{%
108}, 213906 (2012).

\bibitem{circular} D. Leykam, V. V. Konotop, and A. S. Desyatnikov, Opt.
Lett. \textbf{38}, 371 (2013); I. V. Barashenkov, L. Baker, and N. V.
Alexeeva, Phys. Rev. A \textbf{87}, 033819 (2013).

\bibitem{Hadi} J. Pickton and, H. Susanto, Phys. Rev. A \textbf{88}, 063840
(2013).

\bibitem{KPZ} K. Li and P. G. Kevrekidis, Phys. Rev. E \textbf{83}, 066608
(2011); V. V. Konotop, D. E. Pelinovsky, and D. A. Zezyulin, EPL \textbf{100}%
, 56006 (2012); J. D'Ambroise, P. G. Kevrekidis, and S. Lepri, J. Phys. A:
Math. Theor. \textbf{45}, 444012 (2012); K. Li, P. G. Kevrekidis, B. A.
Malomed, and U. G\"{u}nther, \textit{ibid}. \textbf{45}, 444021 (2012); M.
Kreibich, J. Main, H. Cartarius, and G. Wunner, Phys. Rev. A 87, 051601(R)
(2013).

\bibitem{pelinov} P. G. Kevrekidis, D. E. Pelinovsky, D. Y. Tyugin, SIAM J.
Appl. Dyn. Syst. \textbf{12} 1210-36 (2013); see also P.G. Kevrekidis, D.E. Pelinovsky, and D.Y.Tyugin,ÊÊJ. Phys. A: Math. Theor.Ê{\bf 46},Ê365201ÊÊ(2013).

\bibitem{Wunner} H. Cartarius, D. Haag, D. Dast, and G. Wunner, J. Phys. A:
Math. Theor. \textbf{45}, 444008 (2012).

\bibitem{Thaw} T. Mayteevarunyoo, B. A. Malomed, and A. Reoksabutr, Phys.
Rev. E \textbf{88}, 022919 (2013).

\bibitem{AKKZ} F. Kh. Abdullaev, Y. V. Kartashov, V. V. Konotop, and D. A.
Zezyulin, Phys. Rev. A \textbf{83}, 041805(R) {(2011)}; D. A. Zezyulin, Y.
V. Kartashov, V. V. Konotop, Europhys. Lett. \textbf{96}, 64003 (2011).

\bibitem{Canberra} A. E. Miroshnichenko, B. A. Malomed, and Y. S. Kivshar,
Phys. Rev. A \textbf{84}, 012123 (2011).

\bibitem{rudy} M. Duanmu, K. Li, R. L. Horne, P. G. Kevrekidis and N.
Whitaker, Phil. Trans. R. Soc. A \textbf{371}, 20120171 (2013).

\bibitem{combined} Y. He, X. Zhu, D. Mihalache, J. Liu, and Z. Chen, Phys.
Rev. A \textbf{85}, 013831 (2012);
Opt. Commun. \textbf{285}, 3320 (2012).

\bibitem{SA} E. Ott,\textit{\ Chaos in Dynamical Systems} (Cambridge
University Press: Cambridge, 1993).

\bibitem{PhysicaD} B. A. Malomed, Physica D \textbf{29}, 155 (1987); S.
Fauve, O. Thual, Phys. Rev. Lett. \textbf{64}, 282 (1990); W. van Saarloos
and P. C. Hohenberg, Phys. Rev. Lett. \textbf{64}, 749 (1990); V. Hakim, P.
Jakobsen, and Y. Pomeau, Europhys. Lett. 11, \textbf{19} (1990); B. A.
Malomed and A. A. Nepomnyashchy, Phys. Rev. A \textbf{42}, 6009 (1990); P.
Marcq, H. Chat\'{e}, and R. Conte, Physica D \textbf{73}, 305 (1994); T.
Kapitula and B. Sandstede, J. Opt. Soc. Am. B \textbf{15}, 2757 (1998); A.
Komarov, H. Leblond, and F. Sanchez, Phys. Rev. E \textbf{72}, 025604
(2005); J. N. Kutz, SIAM Rev. \textbf{48}, 629 (2006).

\bibitem{Floquet} H. P. Breuer and M. Holthaus, Ann. Phys. \textbf{211}, 249
(1991).

\bibitem{LL} L. D. \ Landau and E. M. Lifshitz, \textit{Mechanics} (Nauka
Publishers: Moscow, 1973).

\bibitem{Sydney} P. L. Chu, B. A. Malomed, G. D. Peng, and I. Skinner, Phys.
Rev. E \textbf{49}, 5763 (1994).

\bibitem{perturbation} T. Tel and M. Gruiz, \textit{Chaotic Dynamics}
(Cambridge University Press: New York, 2006).

\bibitem{Moti} Y. Lumer, Y. Plotnik, M. C. Rechtsman, and M. Segev, Phys.
Rev. Lett. \textbf{111}, 263901 (2013).

\bibitem{Itin} A. P. Itin, R. de la Llave, A. I. Neishtadt and A. A.
Vasiliev, Chaos \textbf{12}, 1043 (2001); Leoncini, A. Neishtadt, and A.
Vasiliev, Phys. Rev. E \textbf{79}, 026213 (2009).

\bibitem{Itin2} A. P. Itin, S. Watanabe, and V. V. Konotop, Phys. Rev. A
\textbf{77}, 043610 (2008); A. P. Itin and S. Watanabe, Phys. Rev. Lett.
\textbf{99}, 223903 (2007).

\bibitem{averaging} V. Zharnitsky, I. Mitkov, and N. Gr\o nbech-Jensen,
Phys. Rev. E \textbf{58}, R52 (1998); F. K. Abdullaev and R. A. Kraenkel,
Phys. Rev. A \textbf{62}, 023613 (2000); A. P. Itin, Phys. Rev. E \textbf{63}%
, 028601 (2001); F. K. Abdullaev, J. G. Caputo, R. A. Kraenkel, and B. A.
Malomed, Phys. Rev. A \textbf{67}, 013605 (2003); F. K. Abdullaev and R.
Galimzyanov, J. Phys. B: At. Mol. Opt. Phys. \textbf{36}, 1099 (2003); A. P.
Itin and A. I. Neishtadt, Phys. Rev. E \textbf{86}, 016206 (2012).
\end{thebibliography}
\end{document}